\documentclass{article}

\usepackage{arxiv}

\usepackage[utf8]{inputenc} 
\usepackage[T1]{fontenc}    
\usepackage{hyperref}       
\usepackage{url}            
\usepackage{booktabs}       
\usepackage{amsfonts}       
\usepackage{nicefrac}       
\usepackage{microtype}      
\usepackage{graphicx}
\usepackage{natbib}
\usepackage{doi}
\usepackage{algorithmic}
\usepackage{algorithm}
\usepackage{amsmath}
\usepackage{amsfonts}
\usepackage{bm}

\title{Online Trading Models with Deep Reinforcement Learning in the Forex Market Considering Transaction Costs}

\author{{\hspace{1mm}Koya Ishikawa}\\
	Department of Industrial Engineering and Economics\\
	Tokyo Institute of Technology, Tokyo, Japan\\
	\texttt{Ishikawa.k.aw@m.titech.ac.jp} \\
	\And
	{\hspace{1mm}Kazuhide Nakata}\\
	Department of Industrial Engineering and Economics\\
	Tokyo Institute of Technology, Tokyo, Japan\\
	\texttt{nakata.k.ac@m.titech.ac.jp} \\
}

\begin{document}
\maketitle

\begin{abstract}
	In recent years, a wide range of investment models have been created using artificial intelligence. Automatic trading by artificial intelligence can expand the range of trading methods, such as by conferring the ability to operate 24 hours a day and the ability to trade with high frequency. Automatic trading can also be expected to trade with more information than is available to humans if it can sufficiently consider past data. In this paper, we propose an investment agent based on a deep reinforcement learning model, which is an artificial intelligence model. The model considers the transaction costs involved in actual trading and creates a framework for trading over a long period of time so that it can make a large profit on a single trade. In doing so, it can maximize the profit while keeping transaction costs low. In addition, in consideration of actual operations, we use online learning so that the system can continue to learn by constantly updating the latest online data instead of learning with static data. This makes it possible to trade in non-stationary financial markets by always incorporating current market trend information.
\end{abstract}


\section{Introduction}
Various machine learning methods are being used as predictive and automated trading models for financial markets. Trading using machine learning provides new methods that were not possible in the past, such as high-frequency trading. In general, machine learning models consider the decision-making process to maximize the return over a specific period of time, and most models are price or trend prediction models. On the other hand, there are two problems with these models: first, they cannot make long-term predictions about transaction costs. In the real market, transaction costs are charged for trading, so to make a profit from trading, one must earn more than the transaction costs, and to maximize the cumulative profit, it is important to earn more profit with as few trades as possible. However, it is very difficult for the existing machine learning methods to make long-term predictions due to their implementation. Another problem is that it is difficult to learn on dynamic data. Machine learning models are often based on the premise that they are trained on static data, and it is not practical to implement them on dynamic data considering the training and other factors.

Here, we have solved the above problem by incorporating reinforcement learning. By incorporating reinforcement learning, it becomes possible to create a model with a long-term perspective and to train the model on the latest data gathered online. The proposed deep reinforcement learning (DRL) model uses a deep learning model as a prediction model for the market and reinforcement learning of trading strategy. In this model, the number of trades decreases as the trading cost increases. In addition, a reward function is set up so that trades can be based on the risk-return of the user’s portfolio and the future market. 

There are several types of financial market, such as stocks, foreign exchange, and virtual currencies, but in this paper, we will mainly consider foreign exchange markets because we can prepare enough data on it. It is assumed that the markets have transaction costs. Transaction costs are incurred when a new position is taken. The investment model we consider assumes that the cost incurred at this time is always constant. It also assumes that users can sell short and that they have large enough assets. This means that they can always trade a fixed amount, regardless of their past earnings. Normally, it would be easier to manage risk if the trading volume could be changed for each transaction, but in this study, it is fixed for simplicity. In Section 2, we introduce related research in deep learning, reinforcement learning, and deep reinforcement learning. Section 3 describes the theory and the details of the proposed model in relation to the deep reinforcement learning model, and Section 4 compares the experimental results of the proposed model and existing methods. Section 5 concludes this paper and outlines remaining challenges.

\section{Related Work}
Reinforcement learning models are often used to solve Markov decision processes. Such models for investment are mainly classified into two types: value function-based and strategy-based. The RRL model\citep{ref1} has been proposed as a strategy-based method, which is a decision-making model that obtains strategies directly from the data and is characterized by the fact that transaction costs are given in the reward function. RRL is a decision-making model that obtains a strategy directly from the data and is characterized by the fact that the reward function, etc., is given a transaction cost, etc. As a value function-based method, it is generally treated as a deep reinforcement learning model and is often used as a decision-making model that selects an action that takes the maximum value based on the value function V, which indicates the expected value of the total reward in the transaction period H, and the action value function Q. In particular, there are many models based on Deep Q-Network\citep{ref2}, and many models combining long short term memory (LSTM)\citep{ref3} and deep neural networks (DNNs) that are especially capable of dealing with time series data have been proposed\citep{ref4}\citep{ref5}\citep{ref6}.Deep learning models that do not use reinforcement learning have also been actively researched. The most popular is the forecasting model using LSTM\citep{ref7}\citep{ref8}, which is well suited to financial markets because LSTM models are time series data. Other models, such as the one in \citep{ref9}, which uses CNNs to learn chart images as features, and one in \citep{ref10}, a restricted Boltzmann machine model, have also been proposed.

\section{PROPOSED MODEL}
\subsection{Definition of transaction costs and Q function}
We chose to use Deep Q-Network (DQN)\citep{ref2} as the deep reinforcement learning model to learn Q functions by approximating them with functions. The Q function is defined as follows.

$$Q(s_t, a_t) =\operatorname{max}\left[\sum^T_{t'=t}\gamma^{t'}r_t'|s_t, a_t\right] $$

where $\gamma$ is the discount rate, and $\gamma$ is the maximum value of the sum of the rewards $r_t$ for T steps when taking action $a_t$ in state $s_t$ at time $t$. In general, the goal of reinforcement learning is to maximize the sum reward $\sum^T_tr_t$   in step T. Therefore, if the Q function can be estimated, the desired strategy function $\pi(s_t)$ is

$$\pi(s_t) = \operatorname{argmax}_aQ(s_t, a).$$

We can express the measure function we want to obtain by transforming the Q function as follows.

$$Q(s_t, a_t) = r_t + \operatorname{max}_aQ(s_{t+1},a).$$

The functionally approximated $Q_\theta(s_t, a_t)$ is learned by the gradient descent method by setting the loss function to

$$Loss = \left( Q_\theta(s_t, a_t) - \left( r_t + \operatorname{max}Q_\theta(s_{t+1}, a) \right) \right) ^2.$$

In the proposed model, we define three types of action $a_t \in \{-1,0,1\}$ positive position, no position, and negative position. Note that when $a_{t}=a_{t+1}=1$, it means that the currency purchased at time t is not purchased additionally at time $t+1$ , but is held without being sold. When $a_{t'}=0$ at time $t' (> t)$ , the currency held is sold, and when $a_{t'}=-1$, the currency held is sold and further margin selling is done. The data is foreign exchange and only the closing price $p_t \in \mathbb{R}$ is used as a feature. We use the price difference $d_t=p_t-p_{t-1}$  as a feature of the model to avoid bias in the data. The input to the model is the price difference $\{d_{t-H-1},\cdots,d_t\}$ from the trade time $t$ to the horizon $H$ steps before and the action $a_{t-1 }$ is connected to the feature $\mathbf{s_t}\in \mathbb{R}^{H+1}$.

The reward function is expressed as

$$R(s_t) = a_t\times d_{t+1} + c|a_{t+1}-a_{t}|$$

where $c \in \mathbb{R}$ represents the transaction cost. The first term of the reward function represents the profit for one step, and the second term represents the transaction cost of the trade. The transaction cost is such that there is no cost if the actions are the same as the previous action. Therefore, it is important for the model to predict the trend and not to change the actions frequently to increase the maximum reward.

The Q function described above is a general definition, but it is not appropriate for this model. What we want to know is not the maximum value of the sum reward, but rather the sum reward of holding a position at time $t$. In other words, if we take the case of a purchase, the following equation is the value we want to know:

$$Q(s_t, buy) = \sum^\infty_{t'=t}\gamma^{t'}r_{t'}(s_{t'},buy).$$

The same is true for selling. By the way, $r(s_t,hold)$ is zero.

If the output of Q defined in this way is $\{Q_{-1},Q_{0},Q_{1}\}$, then $Q_{1}>0$($Q_{-1}>0$), then we can expect to make more than the commission by holding a positive (negative) position. The difference between this model and the conventional deep learning model is that the deep learning model can only see the price change at the next time, while this model considers infinite steps. This allows us to evaluate the price over the long term, which is expected to increase purchase opportunities. By changing the discount rate $\gamma$, we can adjust how much of the long-term reward is considered. In general reinforcement learning models, it is considered to be good for $\gamma$ to be as close to 1 as possible to get close to the global optimal solution, but in the case of financial markets, since the holding period is often finite, 1 is not necessarily a good value and $0.7\sim0.9$ is easier to handle.

\subsection{Online learning using parallel algorithms}
The performance of the existing methods is often evaluated by fitting to existing static data and then using that data for verification. However, in actual operations, the amount of data available online grows. Since the financial market is non-stationary and it is almost obvious that the most recent price fluctuation is more important, in practice it is more important to learn on dynamic data than to learn on static data.

Therefore, we propose a parallel algorithm to keep learning the latest information and to enable online trading at the same time. First, we create two different approximation functions, $Q_\theta$ and $Q_\phi$. Each approximation function is designed to output $Q(s_t)=\mathbf{a_t}=\{a_{-1},a_0,a_1\}$ and the Q value for each action. $Q_\phi$ is the model used for actual trading, and the action with the largest Q value is chosen to be executed in the actual market. $Q_\theta$ is the model for learning behind the scenes. When training, we randomly sample $N$ from the buffer and train the network on the basis of the following loss function:

$$Loss^{tr} = \left(Q(s_t, a_t) - \left(r_t + Q(s_{t+1}, a_t)\right)\right)^2.$$

By defining the Q function as described above, it is expected that the model will learn the latest trends. However, there is a slight problem in determining the data stored in the buffer by the action chosen by $Q_\phi$. This is because the action chosen by $Q_\phi$ is a completely greedy action, and the search for a better optimal solution cannot be sufficiently performed. This is known as the trade-off between search and utilization in reinforcement learning. To solve this problem, the algorithm in this paper does not store in the buffer the data executed by $Q_\phi$, but only the data executed by $Q_\theta$. By making the function $Q_\theta$ behave according to the $\epsilon-greedy$ method, we resolve the trade-off between search and utilization. The $\epsilon-greedy$ method is a search method in which actions are selected completely randomly with probability $\epsilon$ and greedily with probability $1-\epsilon$.

While $Q_\theta$ progressively learns over time, $Q_\phi$ does not. Therefore, we update the performance of the trade model $Q_\phi$ by periodically copying the parameters of $Q_\theta$ to $Q_\phi$. Since it is risky to copy the parameters while holding a position, the timing of copying is limited to when the action $a^{\phi}_t$ selected by $Q_\phi$ is $0$.

\subsection{Overall view}

The function $Q$ is a combination of the LSTM model and a fully connected (FC) model. Since the features input to the model reflect the price fluctuations of the past $H$ steps, we thought that LSTM, which has good compatibility with time series data, would be the best choice. Figure 1 shows the overall view of the proposed model. The state used as input consists of a combination of $Price$ rice and $a_{t-1}$. The features extracted from the LSTM layer are combined with the action $a_{t-1}$ and passed to the full combination layer. The purpose of the all-connected layer is to make it easier to capture the complex features of the market by introducing nonlinear relationships. The total coupling layer consists of two sublayers. In the end, the Q values of the three actions (positive position, no position, and negative position) are output.

\begin{figure}[ht]
\begin{center}
\includegraphics[scale=0.4]{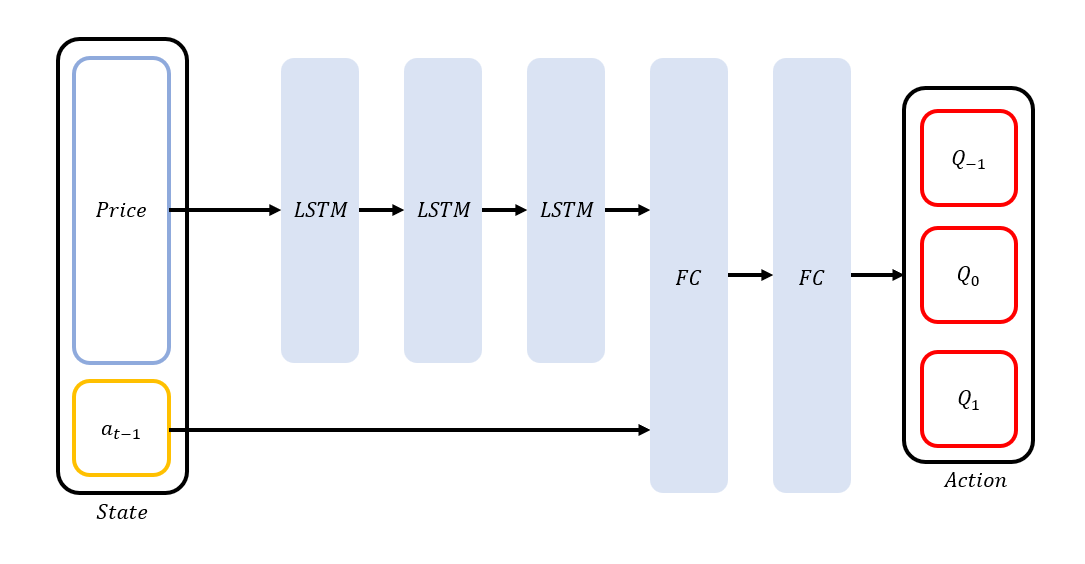}
\caption{Proposed model}
\label{fig:Proposed model}
    \end{center}
\end{figure}

The trading and learning methods of the proposed model are shown in Algorithm 1 and Figure 2. The state $s_t$ is passed to the $Q_\theta$ and $Q_\phi$ models, which each output an action. At this time, $Q_\theta$ is determined according to the $\epsilon-greedy$ method, and $Q_\phi$ greedily selects the action that maximizes the Q value. The action selected by $Q_\phi$ will be used in the trade, and the chosen action for the next state st+1 corresponds to the previous action. The action chosen by $Q_\phi$ is not actually reflected in the trade; rather, the reward $r^\theta_t$ obtained there and the next state $s^\theta_{t+1}$ ($a_t$ contained in this state is the action chosen by $Q_\theta$) together with the data $(s_t,a^\theta_t,r^\theta_t,s^\theta_{t+1})$ are stored in the buffer. The buffer can store $L$ worth of data. If $L$ worth of data has already been stored, the oldest data is deleted and the latest data is stored.

\begin{figure}[ht]
\begin{center}
\includegraphics[scale=0.4]{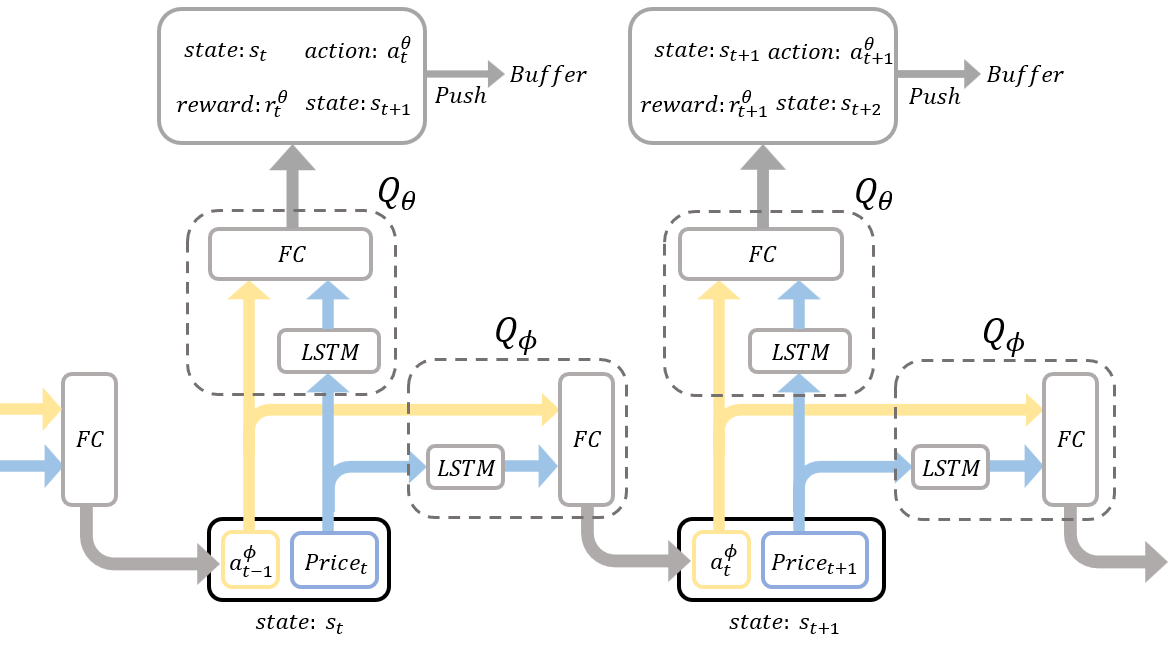}
\caption{Overall view}
\label{fig:Overall view}
\end{center}
\end{figure}

\begin{figure}[ht]
\begin{algorithm}[H]
\caption{Deep Q-Network for online trade}
\label{alg1}
\begin{algorithmic}[1]
\STATE Initialize replay buffer $\mathcal{D}$
\STATE Initialize $Q_\theta$ and $Q_\phi$ function with random weights

\FOR{$t = 1,\infty$}
\STATE Trade according to the action $a^{\phi}_t=\operatorname{max}_a Q(s_t,a;\phi)$ 
\STATE Select $a_t=\operatorname{max}_a Q(s_t,a;\theta)$ with probability $1-\epsilon$
\STATE Execute action $a_t$ in emulator and observe reward $r_t$
\STATE Store transition $(s_t, a_t, r_t, s_{t+1})$ in $\mathcal{D}$
\STATE Calculate $y = r_t + \gamma Q(s_{t+1}, a_t;\theta)$
\STATE Perform a gradient descent step on $(y - Q(s_t,a_t;\theta))^2$
\IF{$a^{tr}_t = 0$}
\STATE Copy parameter $\phi  \leftarrow \theta$
\ENDIF
\ENDFOR

\end{algorithmic}
\end{algorithm}
\end{figure}

\section{NUMERICAL EXPERIMENTS}

\subsection{Data}
The market data used in this paper is FX data from the perspective of large data volume, and the currency pair is USD/JPY. The currency pair used is USD/JPY. The period of use is 1-minute data from 01/01/2020 to 12/31/2020. The reason for selecting the currency pair is due to its high liquidity. In this paper, we use this period as a real dynamic market for online learning. On the other hand, since the existing methods described below are generally not trained online, we use the first half year as training data to train the model, and the second half year as evaluation data. The results in Tables 2 to 5 below are for the period $2020/07/01\sim2020/12/31$.

\subsection{Transaction costs and timestep}
This paper focuses on actual market transaction costs, explicitly giving the costs involved in trading. Not only are transaction costs different for stocks, FX, and virtual currencies, but they also vary depending on the currency pair, time, and brokerage firm used. We will examine several transaction costs, assuming that they are realistic: 0.01\%, 0.02\%, 0.05\%, and 0.1\% of the invested amount. Since the data used in this study has a minimum unit of 1 pip, 0.01\% is the minimum cost in this data. In reality, transaction costs constantly change depending on the liquidity of the market, but for the sake of simplicity, we will assume that transaction costs are always constant.

\subsection{Comparison method}
As a comparison method, we employ a deep learning model using LSTM and LightGBM, which is a gradient boosting tree model. A typical LSTM model is based on the information up to time $t$ and classifies the price increase or decrease at time $t+1$ as a binary value. However, this model does not allow for trades that take transaction costs into account and makes it difficult to compare with the proposed model, so in this paper we use a four-level classification of whether the price increases (decreases) more than the transaction costs. In particular, we use the same four-level classification as the LSTM model.
\newpage
\subsection{Results}

\begin{table}[htbp]
  \centering
  \scalebox{1.2}{
  \begin{tabular}{cccccc}
    cost & 0.01 & 0.02 & 0.05 & 0.08 & 0.1  \\ \hline\hline
    trade num & 5892 & 3290 & 583 & 205 & 245 \\
    return avg & 2.93 & 1.82 & 0.53 & 0.05 & -0.11 \\
    trade length & 7.0 & 6.4 & 7.1 & 8.1 & 9.7 \\
    win rate & 59.5 & 57.3 & 49.2 & 48.1 & 51.9 \\
    shape ratio & 2.04 & 1.77 & 0.68 & 0.15 & -0.40 \\ \hline
  \end{tabular}
  }
  \caption{Performance evaluation by cost}
  \label{tb:Performance evaluation by cost}
\end{table}

Table 1 shows how the performance of the proposed model changes as the cost of the market changes. In the table, trade num corresponds to the number of trades in the evaluation period, and we can see that the number of trades decreases as the cost increases. The return avg shows the average return per trade, which is the percentage increase or decrease in the return of the investment before and after each trade. This is the same as trade num. The trade length is the average number of steps taken per trade. win rate is the number of trades with positive returns divided by the trade num, expressed as a percentage. The shape ratio is the average return divided by the standard deviation of the return, with higher values indicating better performance. Tables 2 to 5 make comparisons with existing models. The value in parentheses for each label is the cost of the experimental environment. In all tables, the number of trades of the comparison method is less than that of the proposed model. This is because the existing model looks for the point where the transaction costs can be exceeded in one step, and this phenomenon is probably caused by the fact that the price rarely changes rapidly in one step. In fact, in an environment where costs are low, the number of trades is at the same level as the proposed method, but as costs increase, the difference tends to become larger. In addition, as the transaction costs rate increases, the number of points at which the variation exceeds the transaction costs rate in a single step decreases. The proposed method, on the other hand, learns whether to exceed the transaction costs in long-term steps rather than in a single step, so these results show that it has more trading opportunities than the comparative methods have. Another feature of the proposed method is that the duration of one trade is much longer than that of the existing methods. The trade length of the proposed method in Table 5 is 9.7, which is long. As mentioned above, it is difficult for the price to change rapidly in a single step, which indicates that the proposed model can learn that it has to hold a position for a long period of time in order to earn more than the transaction costs in an environment where the transaction costs is high. On the other hand, the fact that the compared methods used insufficient trade lengths that allow them to only evaluate in one step shows that it is difficult for them to adapt to environments with high costs.

\begin{table}[htbp]
\begin{center}
\begin{tabular}{c}
  \begin{minipage}[ht]{.50\hsize}
  \begin{center}
  \begin{tabular}{cccc}
    model & DRL & LSTM & LightGBM   \\ \hline\hline
    trade  & 5892 & 5720 & 4889  \\
    return avg & 2.93 & 2.94 & 2.58  \\
    trade length & 7.0 & 3.2 & 2.4 \\
    shape ratio & 2.04 & 1.89 & 0.65  \\ \hline
  \end{tabular}
  \caption{Comparison with existing models(0.01)}
  \label{Comparison with existing models(0.01)}
  \end{center}
  \end{minipage}

\begin{minipage}[ht]{.50\hsize}
  \begin{center}
    \begin{tabular}{cccc}
    model & DRL & LSTM & LightGBM   \\ \hline\hline
    trade num & 3290 & 1701 & 997  \\
    return avg & 1.82 & 1.29 & 0.49  \\
    trade length & 6.4 & 2.3 & 1.7 \\
    shape ratio & 1.77 & 1.31 & 0.43  \\ \hline
    \end{tabular}
  \caption{Comparison with existing models(0.02)}
  \label{Comparison with existing models(0.02)}
  \end{center}
\end{minipage}
\end{tabular}
\end{center}
\end{table}

\begin{table}[ht]
    \begin{center}
        \begin{tabular}{c}
            \begin{minipage}[ht]{.50\hsize}
                \begin{center}
                    \begin{tabular}{cccc}
    model & DRL & LSTM & LightGBM   \\ \hline\hline
    trade  & 583 & 108 & 129  \\
    return avg & 0.53 & 0.42 & 0.51  \\
    trade length & 7.1 & 2.8 & 2.0 \\
    shape ratio & 0.68 & 0.70 & 0.32  \\ \hline
                    \end{tabular}
                    \caption{Comparison with existing models(0.05)}
                    \label{Comparison with existing models(0.05)}
                \end{center}
            \end{minipage}

            \begin{minipage}[ht]{.50\hsize}
                \begin{center}
                    \begin{tabular}{cccc}
    model & DRL & LSTM & LightGBM   \\ \hline\hline
    trade  & 145 & 25 & 20  \\
    return avg & -0.11 & -2.19 & -4.37  \\
    trade length & 9.7 & 1.2 & 2.3 \\
    shape ratio & -0.40 & -1.24 & -1.98  \\ \hline
                    \end{tabular}
                    \caption{Comparison with existing models(0.1)}
                    \label{Comparison with existing models(0.1)}
                \end{center}
            \end{minipage}
        \end{tabular}
    \end{center}
\end{table}

\begin{table}[ht]
    \begin{center}
        \begin{tabular}{c}
            \begin{minipage}[ht]{.50\hsize}
                \begin{center}
                    \begin{tabular}{cccc}
    model & DRL & LSTM & LightGBM   \\ \hline\hline
    trade  & 205 & 20 & 31  \\
    return avg & 0.05 & -1.20 & -2.87  \\
    trade length & 8.1 & 1.5 & 1.2 \\
    shape ratio & 0.15 & -0.49 & -1.99  \\ \hline
                    \end{tabular}
                    \caption{Comparison with existing models(0.08)}
                    \label{Comparison with existing models(0.08)}
                \end{center}
            \end{minipage}
        \end{tabular}
    \end{center}
\end{table}
\newpage
\section{CONCLUSION}

In this paper, we redefine the Q function of the existing DQN model to fit the financial market and incorporate a mechanism that allows constant online learning by devising a buffer. In the existing deep model, the task is to classify and predict the price changes of the next step, and the long-term movements are not paid attention to. However, the proposed model can learn to consider the long-term price changes by learning the sum reward. In actual operations, it is essential to consider transaction costs, and the experiments show that the optimal trading style varies depending on the transaction costs. Their results show that the proposed model is very effective in actual operation.

To improve the accuracy and stability of the model, it is necessary to improve the two aspects of state and action. In our model, only the price of one market is used as the state as the feature value. However, in financial markets, each market does not move independently; rather, multiple markets move in tandem, so it is expected to be effective to incorporate the price movements of multiple markets in the analysis. In addition, not only prices are used as features; economic features can also be incorporated by combining text data from securities reports and sentiment data from social networking services in the case of stock data. In terms of actions, since we traded only a single currency and the trading volume was fixed, we traded at a fixed level. As an improvement, we can consider making it possible to trade at a continuous value without fixing the trading volume and to build a portfolio of multiple currencies. By changing the trading volume depending on the situation, it will be possible to suppress trading in areas with high risk and spend heavily in areas where a large return is expected. Creating a portfolio with multiple currencies is expected to stabilize investment risk by spreading the risk of a single currency across multiple currencies.

In addition, although the market covered in this paper is only that of foreign exchange, we can expect to learn the same thing in the stock market or virtual currency market as well, since the handled data consists of only price fluctuations.

\bibliographystyle{plainnat}
\bibliography{references}

\end{document}